\documentclass{article}

\usepackage{subcaption}
\usepackage[table]{xcolor}
\usepackage{hyperref}
\usepackage{enumitem}
\usepackage{arxiv}
\usepackage[utf8]{inputenc} % allow utf-8 input
\usepackage[T1]{fontenc}    % use 8-bit T1 fonts
\usepackage{hyperref}       % hyperlinks
\usepackage{url}            % simple URL typesetting
\usepackage{booktabs}       % professional-quality tables
\usepackage{amsfonts}       % blackboard math symbols
\usepackage{nicefrac}       % compact symbols for 1/2, etc.
\usepackage{microtype}      % microtypography
\usepackage{lipsum}
\usepackage{graphicx}
\usepackage{academicons}
\usepackage{comment}
\usepackage{color}
\usepackage{multirow}
\usepackage{multicol}
\usepackage{natbib}
\setcitestyle{square}

\title{Android Security using NLP Techniques:\\ A Review}

\author{
Sevil~Sen \\
    WISE Lab., Dept. of Computer Engineering\\ Hacettepe University,\\
    Ankara, TURKEY\\
  \texttt{ssen@cs.hacettepe.edu.tr} \\
   \And
Burcu Can \\
    Research Institute of Information and Language Processing\\
	University of Wolverhampton, \\ Wolverhampton, UK \\
  \texttt{b.can@wlv.ac.uk} \\
}

\begin{document}
\maketitle

\begin{abstract}
Android is among the most targeted platform by attackers. While attackers are improving their techniques, traditional solutions based on static and dynamic analysis have been also evolving. In addition to the application code, Android applications have some metadata that could be useful for security analysis of applications. Unlike traditional application distribution mechanisms, Android applications are distributed centrally in mobile markets. Therefore, beside application packages, such markets contain app information provided by app developers and app users. The availability of such useful textual data together with the advancement in Natural Language Processing (NLP) that is used to process and understand textual data has encouraged researchers to investigate the use of NLP techniques in Android security.  Especially, security solutions based on NLP have accelerated in the last 5 years and proven to be useful. This study reviews these proposals and aim to explore possible research directions for future studies by presenting state-of-the-art in this domain. We mainly focus on NLP-based solutions under four categories: description-to-behaviour fidelity, description generation, privacy and malware detection.
\end{abstract}

%\begin{keyword}
%Keywords: semantic segmentation, deep convolutional neural %networks, fully convolutional networks, survey
%\end{keyword}

\section{Introduction}
\label{intro}

With the advancement of mobile devices and communication technology, mobile devices have become an integral part of our lives. They provide many useful functions such as the ability to read/write e-mails, indicate nearby facilities, video conferencing %, and voice recognition, 
to name but a few. According to \citep{ciscoReport}, 5.6 billion people will have mobile devices by 2020, which is more than the number of people that have electricity (5.3 billion), running water (3.5 billion) and cars (2.8 billion). Furthermore, there have been reported almost 2.5 million available applications (apps) on Android official market (Google Play) %, and almost 2 million available applications on iOS official market (Apple app store) 
in the second quarter of 2019 \citep{statista}. However, the popularity and adoption of mobile devices and mobile apps also attract attackers in order to harm mobile devices, and steal private information of mobile users. We still see that mobile malware has continued to increase in scope and complexity \citep{mcafee2019}. 

Unlike traditional application distribution mechanisms, mobile apps are distributed centrally, so mobile markets beside application package contain application metadata such as application definitions, user scores and reviews. Having a central app deployment mechanism not only has enriched studies on software engineering, but also has potential in augmenting security approaches by using non-technical data available in app stores \citep{sesurvey:2016}. For example, while application descriptions could give an idea about the intention of application developers, user reviews could give some direct information about the experience of users on applications. Even application codes could be considered as a form of textual data. Natural language processing (NLP) provides a way for understanding and using such textual data for various reasons including security. It is a new application area of NLP that is accelerated mainly with the availability of useful metadata on application stores. The main aim of this study is to review such studies that integrate NLP with Android security, by utilizing any type of textual data provided with apps. %Such studies on mobile security has accelerated since 2015. 
With this study, these trends are outlined and future research directions in this domain are discussed. 

In this study, the use of NLP in Android security is explored under four categories: description-to-behaviour fidelity, description generation, privacy, and malware detection. %From the privacy and security point of view, if the functionality of applications is given in sufficient detail in their descriptions, then the requirement of requested permissions could be well understood. This is defined as description-to-permission fidelity in the literature \citep{Autocog}. 
\textit{Description-to-behaviour fidelity} part reviews studies that aim to discover inconsistencies between app behaviour and metadata. %, namely description and user reviews. 
In \textit{description generation} part, %\color{red}generation-based research\color{black}, 
studies that aim to generate app descriptions, privacy policies, explanations of malware, or any other \color{black} metadata, particularly security or privacy \color{black} related text generations are analyzed. \textit{Privacy} part deals with studies focusing on discovering sensitive user inputs \color{black} that could help detecting data leakage in apps. Finally, \textit{malware detection} part explores the usage of metadata on malware analysis and detection. %The keywords of the study can be seen in Figure \ref{fig:wordcloud}. As it is seen in the word cloud, app descriptions come forward as one of the most applied metadata in the proposed applications. 

All of these research problems analyzed in this article address either human generated textual data or aim to generate textual data according to the natural language rules. Therefore, using the methods to understand natural languages, namely NLP, enables also tackling with the description-to-behaviour fidelity problem or malware detection by using any metadata of apps that are written in a natural language. 
Security-related research could consider the problem as a keyword searching problem \citep{Whyper, supor:2015}. For example, a keyword search performed in an application's description can predict that there is no sufficient detail in the description regarding the READ\_CONTACTS permission. However, the absence of a keyword does not necessarily imply that such meaning does not exist in the description. There could be semantically related other words that could lead to a similar meaning and if we consider the vocabulary coverage of a natural language, we can conclude that it is nearly impossible to cover all the intended keywords. Meaning is yet more than lexical meaning that consider's only the meaning of a single word. Understanding is a lot more contextual than this. Therefore, most of the tasks require more natural language understanding rather than just searching for some keywords. Some studies \citep{Autocog, AC-Net:2019} have already shown that involving more semantics in security-related problems also enhances the success of the studies. Semantics is only one dimension in NLP and other dimensions (such as syntax) that enables to understand a language will be also reviewed in the following section (see Section \ref{background}).  
\color{black}

The paper is organized as follows. The next section introduces the basic concepts in Android security and NLP. The current state-of-the-art on using NLP for Android security is summarized in four sections, namely description-to-permission fidelity, description generation, privacy and malware detection in Section \ref{sec:dpf}-\ref{sec:md}. Each section separately includes discussion of the related studies for addressing corresponding problem and explores possible future research directions. Section \ref{sec:RW} shortly gives traditional approaches in Android security. Section \ref{sec:disc} discusses how these approaches could be complemented with the NLP-based approaches reviewed in this study. Moreover, the strengths and weaknesses of the proposed approaches from the NLP point of view are discussed in this section, and the possible future directions for researchers are presented. Finally, Section \ref{sec:Conc} includes concluding remarks.

\section{Background} 
\label{background}

In this section, background information on Android security, particularly permissions mechanism and  NLP will be covered. Android, which is based on a modified version of the Linux kernel, is still among the most targeted platforms by attackers. Android security is mostly related to protect mobile devices and users from harmful activities carried out through apps. One of the basic Android security mechanisms is isolating these app resources by assigning each app a unique user ID and running it in a sandbox. Another important mechanism is to control access to user sensitive data and to certain system resources by using permissions, which is described in details below. Android has been modifying its architecture to improve security, but that is only beneficial to users who download the latest version of Android, which is rarely the case \citep{symantec2017}.

\subsection{Permissions}

Permissions are one of the key points of Android security mechanism. The permissions required by app must be listed in application's manifest file. According to their protection level, Android permissions are generally divided into two groups: normal and dangerous permissions. Since normal permissions cover access areas of an application that are outside the app's sandbox \citep{permOverview}, they do not need to be granted by users. On the other hand, users have to grant explicitly dangerous permissions because such permissions request to access users' sensitive data such as contact lists, call logs or certain system features such as camera, microphone \citep{permOverview}. The way Android asks users to grant dangerous permissions has changed with Android 6.0. Before Android 6.0, all dangerous permissions had to be granted at the installation time. In Android 6.0 and in the higher versions, users are asked to grant dangerous permissions at runtime. 

Dangerous permissions in Android are listed in Table \ref{tab:perms}. Users expect explanations for the usage of such permissions in app descriptions or privacy policies, since such permissions could affect the privacy of users or the normal operation of the system \citep{permOverview}. This is called description-to-permission fidelity in the literature \citep{Autocog}. Since permissions are one of the important aspects of Android security and, dangerous permissions are expected to be explained in app metadata, the NLP studies on Android security are mainly shaped around permissions.

\begin{table}[t]
\scriptsize
\caption{Dangerous Permissions \citep{perms}}
\begin{center}
\begin{tabular}{| c | l |}
    \hline
    \textbf{Group} & \textbf{Permissions} \\
    \hline
    \multirow{1}{*}{Calendar} & READ\_CALENDAR\\
                                & WRITE\_CALENDAR \\\hline
    \multirow{1}{*}{Camera} & CAMERA \\\hline
    \multirow{1}{*}{Contacts} &  READ\_CONTACTS\\
                                & WRITE\_CONTACTS\\
                                & GET\_ACCOUNTS\\\hline
    \multirow{1}{*}{Location} & ACCESS\_BACKGROUND\_LOCATION\\
                            & ACCESS\_FINE\_LOCATION \\
                            & ACCESS\_COARSE\_LOCATION  \\
                            & ACCESS\_MEDIA\_LOCATION  \\\hline
    \multirow{1}{*}{Microphone} & RECORD\_AUDIO  \\\hline
    \multirow{1}{*}{Phone} & READ\_PHONE\_NUMBERS \\
                        & READ\_PHONE\_STATE \\
                        & CALL\_PHONE\\
                        & READ\_CALL\_LOG\\
                        & WRITE\_CALL\_LOG \\
                        & ADD\_VOICEMAIL \\
                        & USE\_SIP \\
                        & PROCESS\_OUTGOING\_CALLS \\
                        & ANSWER\_PHONE\_CALLS\\
                        & ACCEPT\_HANDOVER \\\hline
    \multirow{1}{*}{Sensors} & BODY\_SENSORS \\
                            & ACTIVITY\_RECOGNITION\\\hline
    \multirow{1}{*}{SMS} &  SEND\_SMS \\
                    & RECEIVE\_SMS \\
                    & READ\_SMS\\
                    & RECEIVE\_WAP\_PUSH\\
                    & RECEIVE\_MMS\\\hline
    \multirow{1}{*}{Storage} & READ\_EXTERNAL\_STORAGE \\ 
    & WRITE\_EXTERNAL\_STORAGE \\
    \hline
\end{tabular}
\end{center}
\label{tab:perms}
\end{table}

\subsection{NLP}

Natural language processing (NLP) is concerned with understanding human languages through computer programs and therefore addresses interactions between humans and computers. %It is both a subfield of linguistics and computer science. 
NLP involves tasks from the linguistic levels of a language, such as part-of-speech (PoS) tagging, stemming, parsing, and semantics to better understand the natural languages. %, and on the other side it involves problems that are more \textcolor{blue}{complex?} in the application level, such as machine translation, text summarization, named entity recognition etc.
Here, we will not mention about all the tasks in NLP, but  cover some of them, which are frequently utilized in the security or privacy fields, therefore we aim to ease reading of the article for the readers who are outside the NLP domain. 

\begin{figure}[t]
\centerline{\includegraphics[scale=0.7]{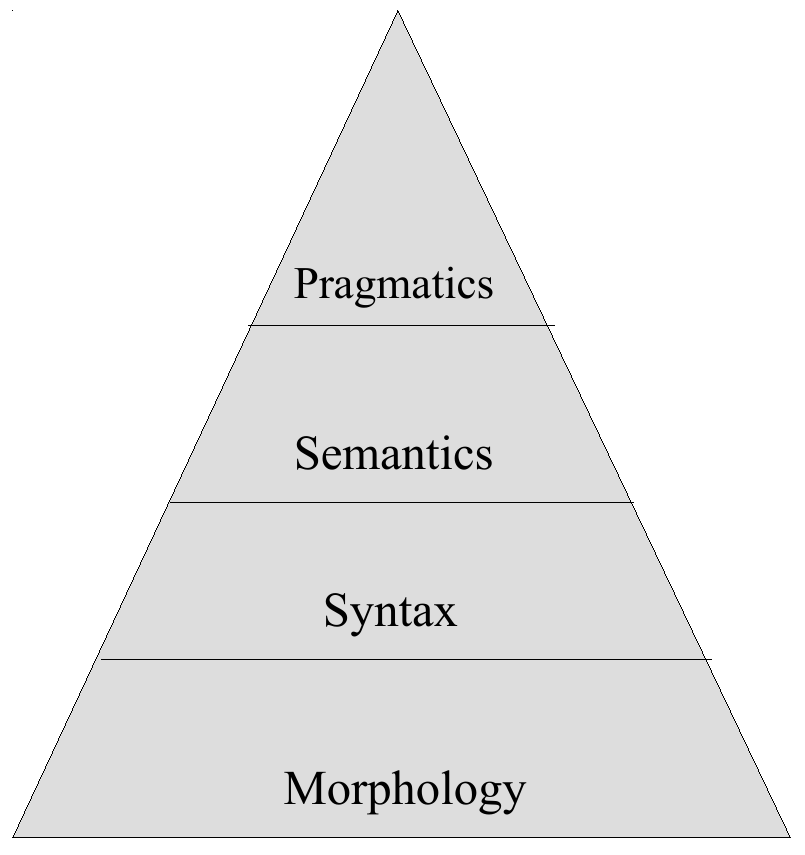}}
\caption{NLP Pyramid}
\label{fig:pyramid}
\end{figure}

There are several linguistic levels of a natural language (see the NLP pyramid in Figure \ref{fig:pyramid}). The fundamental levels are syntax and semantics. 
%such as morphology, syntax, and semantics (other levels such as phonetics or pragmatics are out of the security domain). 
Syntax defines the rules of a language that describe the word order in a given sentence and the functionalities of the words in that sentence. Part-of-speech (PoS) tagging is the task of assigning a syntactic category to each word in a given sentence depending on its syntactic role in that context, such as a noun, an adjective. For example, it could be crucial to determine the meaning of \textit{contact} depending on its PoS tag (i.e. if it is a noun then it possibly refers to violations in regard to contact lists). Dependency parsing gives a more detailed syntactic structure of a sentence along with the relationships between the words in that sentence. For example, \textit{secure} is a modifier and \textit{applications} is the head of this modifier in the noun phrase \textit{secure applications}, which is typically the subject or the object of a verb in the sentence leading to a verb phrase, such as \textit{installing secure applications}. Dependency parsing is one of the frequently utilized tasks in security and privacy. For example, in the description-to-permission fidelity problem, permissions are usually verb phrases and need to be extracted from descriptions or reviews automatically
\citep{Autocog}\citep{SmartPi:2019}, or sensitive inputs are usually noun phrases in privacy policies \citep{enhancingdtob:2017}. 

%Morphology is concerned with the smallest meaning bearing units that are called morphemes in a word. For example, the word \textit{bookings} is made up of the following morphemes: \textit{book}, \textit{ing}, \textit{s}. Morphological analsysis involves extracting those morphemes out of each word and also assigning a syntactic category to each morpheme.
Language is usually sparse due to different and usually informal writing styles of different users (i.e. user reviews), or due to different forms of the same word (having inflectional/derivational morphemes at the end of the word) such as \textit{contacting} and \textit{contacts}. Those differences are standardized to reduce the number of features extracted from data. Stemming is a common way to 
reduce word forms into a single form by filtering out inflectional morphemes from the end of the words, which is related with the morphology level (Figure \ref{fig:pyramid}). 

%For many tasks that deal with the meaning of the word, stemming is required to reduce the sparsity in data because every inflected form of a single word would have the same meaning, and therefore they need to be considered as a single token in a given task. 
Semantics is concerned with the meaning of a word in a given context and it is usually coupled with syntax. For example, the meaning of the word \textit{contact} is only obvious, if we know that it is either a verb or a noun. Therefore, most semantic related problems (i.e. permission extraction from  descriptions or detection of sensitive words in a given user input) involve also syntactic parsing or PoS tagging as a preprocessing task. Meaning can be represented in different forms. Lexical dictionaries such as \citep{Wordnet} for English are very common to obtain semantic information of a word through its synonyms and hypernyms/hyponyms\footnote{A hyponym's meaning is involved in a hypernym's meaning. \textit{Contact} can be considered as the hypernym of \textit{address} and vice versa.}. WordNet has been used in various problems such as description-to-permission fidelity, or privacy for especially finding the synonyms of words. Recently, the concept of neural word embeddings has risen with the neural model called word2vec \citep{word2vec} where the representations (that involve semantics of words) of words in similar contexts (thereby having similar meanings) are urged to be located closer to each other in the vector space, thereby having similar word embeddings. Word2vec \citep{word2vec} is based on a two layer neural network where the hidden layer weights are extracted to be used for the word representations.  Word2vec \citep{word2vec} has also been used in description-to-behaviour fidelity \citep{AC-Net:2019, SmartPi:2019}, description generation \citep{permReqRec:2018}, privacy \citep{findingClues:2018} for assessing the similarities between the meanings of words/phrases, and malware detection (e.g. apk2vec \citep{apk2vec:2018} adopts the same architecture as word2vec where the app representations are obtained)\color{black}. Moreover, some studies \citep{slavin2016, guileak:2018, featuresmith:2016, enhancingdtob:2017} build their ontologies to learn the semantic relations between concepts.

The top level in the NLP pyramid corresponds to pragmatics, which handles the context in the text as a whole by again concerning with the semantics with a deeper context. To our knowledge, this level has not been studied under Android security yet. \color{black}

\color{black}

\section{Description-to-behaviour Fidelity}
\label{sec:dpf}

One of the important usages of app metadata is to discover inconsistencies between app behaviour and app metadata. For instance, if the functionality of apps is given in sufficient detail in their descriptions, then the requirement of requested dangerous permissions could be well understood. This is defined as description-to-permission fidelity \citep{Autocog}. From the security point of view, the lack of an explanation about the requirement of a dangerous permission in app's description creates a suspicion. Such suspicious apps could then be subject to analysis by more resource-intensive techniques. It is a new application analysis approach that could complement traditional analysis techniques. Moreover, the proposed approach could assist users before installing apps or app developer could use it in order to improve their descriptions and create user-understandable descriptions. Lastly, it could be used the other way around by generating semantic permissions based on descriptions \citep{ASPG:2014}, if descriptions well represent the functionality of an app.

Although most studies in the literature focus on finding semantic relatedness between permissions and descriptions, recent advancements employ also user reviews and privacy policies for extracting useful information. App descriptions have a limited space, so user reviews and privacy policies could consist of more information about security and privacy of applications. For instance, while user reviews could give information about the positive/negative experience of users about dangerous permissions, privacy policies focus more on the usage of sensitive and private data whose access requires dangerous permissions. Moreover, since privacy policies generally contain simple sentence structures due to developers being advised to do so, NLP techniques are promising on such texts. Moreover, approaches for finding inconsistencies in privacy policies are recently proposed and explored here. %While user reviews could give information about the positive and negative comments on the usage of a dangerous permissions,  

To sum up, in this study, approaches on description-to-behaviour fidelity are covered under the following four sections: descriptions, user reviews, privacy policy, and static/dynamic analysis. Most of the studies focus on metadata for understanding the purpose of a permission. However, the reason of an app to access sensitive data could also be obtained by analyzing the app's code or runtime behaviours. Therefore static and dynamic analysis techniques combined with NLP-based techniques for this purpose are also included. % in this section.  

\subsection{Descriptions}

Permissions are used by apps for a variety of reasons. However, while the usage of a set of dangerous permissions could be understandable for an anti-virus solution, the usage of the same set of permissions in a daily horoscope app could be an indication of malicious intention. For such apps which use dangerous permissions, the developer is expected to state the requirement of dangerous permissions in its description. While  anti-virus apps mentioned above are expected to do so, the malicious horoscope app could hide that information from its description. Even though, users are asked explicitly to grant dangerous permissions during installation or runtime, most users are not conscious or aware of such requests \citep{Felt:2012}, so app
descriptions could be a more preferable choice for naive users. This is the main motivation behind the studies below, so they propose automatic (fully or half-fully) approaches for finding inconsistencies between descriptions and requested permissions.

The first study on assessing description-to-permission fidelity is carried out by \citep{Whyper}. The framework called W\texttt{HYPER} (WHP PERmissions) uses NLP in order to identify sentences that describe the need for requested permissions and, compares the proposed approach with a keyword-based approach. In the keyword-based method, a word can have different meanings in different sentences. On the other hand, since W\texttt{HYPER} proposes a semantic-based approach, it performs much better than the keyword-based approach. At first, W\texttt{HYPER} converts each sentence into their First-Order Logic (FOL) representations after pre\-processing the sentences (detecting sentence boundaries, handling abbreviations, etc.) in app definitions. Then, it creates semantic graphs of permissions from application programming interface (API) documents, and checks whether these semantic graphs associate with description sentences or not. The success of the proposed framework depends on being sufficiently comprehensive of semantic graphs of the permissions.

While W\texttt{HYPER} is proposed as a means to alleviate the shortcomings of a keyword-based approach (confounding effects that a word can have different meanings and semantic interference that describes the usage of a permission without using a particular word), \citep{ACODE:2015} proposes a keyword-based approach called A\texttt{CODE} due to being lightweight for large datasets, because A\texttt{CODE} does not require labelling app descriptions. \color{black}By combining static analysis and text analysis, A\texttt{CODE} performs better than the keyword-based approach used for comparison in W\texttt{HYPER} \citep{Whyper} and produces comparable results with W\texttt{HYPER}. Relevance weights inspired from the information retrieval field are used for extracting the keywords, which are used to match queries with relevant documents. In this study, due to the lack of labelled dataset, descriptions that involve permission statements are regarded as relevant documents, and the rest is regarded as irrelevant documents. Keyword extraction is performed  by ranking the words based on the relevance weights, and furthermore combining the results with the code analysis. \color{black} Unlike other studies in the literature, it can be applied to different languages without much effort and change. %\color{red}Moreover, the study involves experiments on Chinese dataset along with the experiments on English dataset. Chinese requires one more NLP step for word segmentation, since there is no boundary between words in Chinese language. \color{black}

A\texttt{UTO}C\texttt{OG} also employs NLP techniques in order to relate descriptions with permissions \citep{Autocog}. Since API documents are not sufficient for having the complete semantic patterns of some permissions, only descriptions are employed in order to extract semantic information in A\texttt{UTO}C\texttt{OG}. Semantic relatedness between descriptions and permissions is measured using Explicit Semantic Analysis (ESA). Instead of using a dictionary based corpus like WordNet \citep{Wordnet} as done in W\texttt{HYPER}, ESA uses a large knowledge base (i.e., Wikipedia) in order to create vectorial representations of the text. A\texttt{UTO}C\texttt{OG} extracts noun phrases from descriptions and creates a semantic relatedness score matrix among them using ESA. Frequent noun phrases are grouped together and a relatedness dictionary is created between the noun phrases (np). To relate permissions with noun phrases, correlation between noun phrases and permissions are defined using a threshold value. Those noun phrases are extended by pairing them with related np-counterparts. Therefore, each permission has a list of related noun phrases and np-counterpart pairs. Given a description, A\texttt{UTO}C\texttt{OG} extracts noun phrases and np-counterpart pairs and then identifies whether the need for a permission is stated in the description or not. While A\texttt{UTO}C\texttt{OG} is a fully-automated approach and performs much better than W\texttt{HYPER}, it could extract semantic relationships that may not actually exist, which may lead to false positives.

%The framework called A\texttt{C}-N\texttt{et} utilizes recurrent neural networks (RNNs) in order to learn and detect semantic relations. Labeled descriptions for 11 different permission groups are used for training and predictions for learned permissions are generated as probability distributions in the model. Since RNNs are able to remember previous inputs, they are good at modelling sequential data. However, RNNs suffer from vanishing gradient problem due to using the back-propagation method for learning. This problem is defined as short-term memory problem in RNNs. Gated Recurrent Unit (GRU)~\citep{Kyunghyun:2014:GRU} as used in A\texttt{C}-N\texttt{et}, offers a solution to RNNs' short-term memory problem by employing gates in order to adjust information flow in network. 

In AC-NET (Assessing Consistency based on Neural Network) \citep{AC-Net:2019}, %proposes another method to assess the consistency between descriptions and permissions. 
unlike Whyper \citep{Whyper} and Autocog \citep{Autocog}, semantic representations of words are obtained from word embeddings. % that represent words in a low dimensional continuous vector space. Word embeddings have shown superiour performance in many NLP tasks including machine translation, question answering, text summarization etc.  
Word2vec \citep{word2vec} is trained on Android app descriptions collected from Google Play. Therefore, the word embeddings used in AC-NET have domain-specific features. Once the word embeddings are learned through the word2vec model, Recurrent Neural Networks (RNNs) \citep{LSTM} are used for generating the compositional meaning of a given decsription by processing each word at a time and gives the overall meaning (i.e. vectorial representation) of the given sentence in a description. Compositional semantics states that the meaning of a phrase or a sentence is obtained from the composition of the meaning of each word in a sequence. In AC-NET, Gated Recurrent Unit (GRU) \citep{GRU}, a type of recurrent neural network, is used for the vanishing gradient problem in the standard recurrent neural networks for the long sequences. %Long Short Term Memory Network \citep{LSTM} is also another type of recurrent neural network that is used to solve the gradient problem \citep{Kabukcu}, but in this work GRUs are preferred because of their simpler architecture compared to LSTMs. 
The word2vec trained embedding of each word in a description is is fed through a bidirectional GRU, where one GRU processes the sentence in the forward order and another GRU processes the sentence in the reverse order, and finally the outputs of each GRU is concatenated to have a compositional representation of the given sentence. A probability of the given sentence implying the given permission is obtained by applying a sigmoid function on the output representation of the sentence. The current state-of-the-art results are obtained from this model due to the performance of both using domain-specific word embeddings and recurrent neural networks. In a very recent study called DesRe \citep{alecakir2021attention}, two models that use also GRU are proposed. Here, the attention mechanism is also employed for the problem for the first time in the literature. While the first model identifies permission sentences in a description, the document-based model based on hierarchical attention network represents an app description as a whole differently from other studies in the literature. By using the attention mechanism, the permission statement words or sentences in app descriptions are found and shown to have positive effects on the results. Moreover, the study introduces a new dataset called DesRe \citep{DesRe}, which includes app descriptions and five user reviews declared to be most helpful by other users for each application.
\color{black}

\subsection{User Reviews}

AutoReb \citep{autoreb:2015} introduces the concept of  \textit{review-to-behaviour fidelity} in Android apps as the first work that explores employing the user review information to detect any security related behaviour of an app. It is a keyword-based approach where the keywords are manually extracted by analyzing the co-occurrence of the words with the initially defined two-word set (\{\textit{security, privacy}\}). Similar to Whyper \citep{Whyper}, an information retrieval method called query expansion is used to find the relevant reviews that include security-related content. Once the reviews are expanded, Bag of Words (BOW) features are extracted from those reviews to be used in classification training. Spare linear Support Vector Machine (SVM) is used for training the model for the multi-label classification to assign a given review to one of the following security behaviour categories: \textcolor{black}{spamming, financial issue, over-privileged permissions, and data leakage}. Once the security behaviour is annotated using the query expansion method, a crowdsourcing method is used to finalize the label of each app because there could be conflicts among different reviews written for the same app. The labelling provided by the SVM classifier for each review is regarded as crowdsourcing opinions gathered from different users. Expectation Maximization algorithm (EM) is used along with Maximum Likelihood Estimate (MLE) to aggregate the labels obtained from the classifier with the application level behaviour, thereby making a binary prediction for each label. The study proves that user reviews play an important role in analyzing the application behaviours, although the user expectations could be different and the corresponding reviews are quite personal.\color{black}

A very recent study \citep{SmartPi:2019} proposed to find permission indications from user reviews, differently from the  previous studies. The framework called SmartPI includes four phases: data collection, reviews selection, reviews clustering, and permission inference. In the data collection phase, apps, app descriptions, permission docs, API docs and user reviews are collected. Representative words of permissions are extracted by using these documents. Firstly, noun or noun phrases and word pairs $\langle$verb, noun$\rangle$ are extracted from permission documents and API documents, then these representative words are enhanced in two ways. In the first way, synonyms of those words are extracted from WordNet \citep{Wordnet}. Secondly, co-occurence words of representative words are extracted from descriptions according to their co-occurrence frequencies. In review selection phase, both user reviews and permission-representative words are formed as feature vectors using word2vec \citep{word2vec}. Then, each review's similarity to the permission-representative words are calculated by using cosine similarity. According to the estimated similarities, funtionality-relevant user reviews are selected. Reviews are grouped into 10 (number of permissions) clusters using Biterm Topic Model (BTM) \citep{BTM:2014}. Each cluster represents a permission. Finally, a review is mapped to a cluster in the permission inference phase. The proposed study is compared with A\texttt{UTO}C\texttt{OG} \citep{Autocog}, and shows slightly better results. Furthermore, two approaches are evaluated on 200 apps randomly collected from Google Play. This test supports the hypothesis of the study that user reviews are more representative than app descriptions.% in order to indicate the usage of permissions. 

Another recent study \citep{alecakir2020discovering} also employs the user reviews for the description-to-permission fidelity problem. In that study, the most helpful reviews (rated as five stars) are involved in detecting any inconsistency between descriptions and the application behaviour. The model is trained by using the annotated description dataset called DesRe \citep{DesRe}, and it incorporates the reviews during testing. To this end, either the description or both the description and reviews are used in testing. If the description is sufficient to claim that there is not such an inconsistency between the description and the permission, and the permission is already mentioned in the description, then the reviews are not used. However, in the lack of a strong claim for such an inconsistency, the reviews are employed along with the descriptions in testing. The model is based on an RNN structure similar to \citep{AC-Net:2019}, but uses Long Short Term Memory Network \citep{LSTM}, which is also another type of recurrent neural network. The results show that using reviews improves the detection capability of the model for particular permissions. \color{black}

Last but not least, \citep{shortText:2019} analyze the user reviews, but unlike the previous work the authors try to find the connections between security and privacy-related reviews (SPR) and security and privacy-related app updates (SPU). Once the reviews are collected from Google Play, a keyword list is generated half manually, and SVM is used as a classifier to categorize the reviews either privacy, security-related or not by employing the character n-grams of words in the reviews that are already stemmed. An app's updates and relevant security or privacy-related changes are obtained by static code analysis on different versions of the app that are released on the date of reviews. Finally, SPR and SPU belonging to different versions of the app are mapped. The results show that majority of the SPR is a predictor for SPU.  \color{black}

\subsection{Privacy Policy} 

Personal and sensitive user data is defined but not limited to by Google as ``personally identifiable information, financial and payment information, authentication information, phonebook, contacts, SMS and call related data, microphone and camera sensor data, and sensitive device or usage data'' \citep{GooglePrivacy:2017}. Google expects developers to be transparent about disclosing the collection, use, and sharing of such personal and sensitive data, and limiting the use of the data to the purposes disclosed, and the consent provided by the user \citep{GooglePrivacy:2017}. Such information is expected to be given in privacy policies and Google announced that it is going to remove applications which do not comply with Google's User Data Policy starting from March, 15 2017 \citep{aboutPrivPolicy:2017}. This privacy and security enforcement by Google and the character limitation on app descriptions make the usage of privacy policies eligible for Android security. Moreover, the guidelines for developers suggest using a simple language in privacy policies in order to be understandable by various users. Hence, such documents could be easily processed by NLP techniques. For example, a recent study is able to extract general sentence structures in Android privacy policies successfully \citep{enhancingdtob:2017} and verbs in these structures conform to common privacy policy keywords \citep{policyKeywords}. 

%This section especially focuses on the usage of privacy policies for enhancing description-to-behaviour fidelity. However as shown in subsequent sections, privacy policies are used for other purposes in Android security such as automatic generation of policies that explain the usage of sensitive data in application codes, malware detection. 

%\textcolor{blue}{ Ayrica privacy policy ile literaturdeki genel  calismalara referans verilebilir:  7.3 in \citep{enhancingdtob:2017} }

Some recent studies explore the use of privacy policy for enhancing the description-to-behaviour fidelity \citep{enhancingdtob:2017, revisitingdtob:2016}. TAPVerifier \citep{enhancingdtob:2017} claims that descriptions are not sufficient for assessing the description-to-behaviour fidelity, since developers might not declare all usages of sensitive data in app descriptions that have a character limit. Privacy policies could give more information about that. The new update on the User Data Policy of Google, which forces developers to provide privacy policy \citep{aboutPrivPolicy:2017}, supports this hypothesis. TAPVerifier employs a two-stage analysis in order to complement other studies such as Autocog \citep{Autocog} and decreases their false positives: privacy policy analysis, code and permission analysis. While privacy policy analysis extracts the necessity of a requested permission from the app's privacy policy, code analysis extracts the permissions used in the code by using PScout \citep{pscout:2012}, which maps API calls with permissions. The permissions of the third party libraries are also included in this study as one of the enhancements of the previous study of the same authors \citep{revisitingdtob:2016}. 

The privacy policy analysis of TAPVerifier \citep{enhancingdtob:2017} largely relies on NLP techniques. After preprocessing and splitting privacy policies into sentences by using \citep{NLTK}, the dependency parse trees of those sentences are generated by using the Stanford Parser \citep{StanfordParser} along with the PoS tags of the words within each sentence\color{black}. Then sentences are mapped to  one of the 9 manually defined \color{black}semantic patterns. These are structure(s) of the actions that need to be mentioned in a privacy policy such as data collection, data access, data disclose. If a sentence cannot map to any semantic pattern, it is removed. Then, resources in the extracted sentences are determined by extracting noun phrases in the unconditional clauses. Co-reference resolution\footnote{Co-reference resolution is the task of detecting expressions in a sentence that refer to the same entity.} is also applied to find the entities that the pronouns point to, which shows that it improves the overall performance.  \color{black} If the sentence that the resources are referenced in is a negative one, in other words if it contains negative words such as \textit{no}, \textit{unable}, \textit{hardly}, then such resources are removed, since it means that those resources are not referenced in any action in the app. Finally, the resources are mapped to resources of permissions, which are defined by analyzing the permission's description and the documents of APIs that are matched by using PScout \citep{pscout:2012}. If the similarity of resources to the resources of permissions, which is calculated by using ESA \citep{ESA} a method that gives a vectorial representation of text on WordNet \color{black} exceeds a predefined threshold, it means that the privacy policy includes an indication of the requirement of that permission. Please note that TAPVerifier also analyzes the privacy policies of the third party libraries.

PPChecker \citep{trustpolicy:2016} uses  NLP for identifying the problems in privacy policy, such as incomplete privacy policies that lack the details related to the app behaviours, incorrect privacy policies that conflict with the app behaviour, and inconsistent privacy policies that conflict with the third party libs' privacy policies. The authors use the pre-defined verb categories that are used in privacy policies such as \textit{collect}, \textit{use}, \textit{retain}, and \textit{disclose} verb groups \citep{eddy, Breaux}. Similar to TAPVerifier \citep{enhancingdtob:2017}, text is first split into sentences using NLTK and sentences are parsed by Stanford Parser \citep{stanford-parser} to extract the dependency relations between the phrases in each sentence. A bootstrapping mechanism is used to automatically find any pattern in the privacy policies by giving the seed pattern SVO (subject-verb-object) initially. All the matched sentences with this pattern are collected and the subjects and objects in those sentences are also collected. Then new patterns are extracted by using the collected subjects and objects in the dependency trees. Therefore, this study improves upon TAPVerifier \citep{enhancingdtob:2017} by automatically generating the patterns. The patterns are ranked by using the number of sentences that can match the pattern. Finally, sentences are selected by using the generated patterns in the previous step. Similar to TAPVerifier \citep{enhancingdtob:2017}, negation analysis is also performed to detect the negative sentences. However, in this work a negative word list is also employed. Information elements are extracted from the selected sentences that include the main verb, subject, and the object. A constraint is also used to remove the sentences that are conditional on an action, such as \textit{if you ...}. The system performs static code analysis and description analysis by using AutoCog \citep{Autocog}. The model gives a high 
precision of detecting suspicious privacy policies. \textcolor{black}{It also discovers inconsistencies between descriptions and privacy policies from the point of permission view.}

%\textcolor{blue}{Bu paragrafi cikarsak mi, Android policylere bakmiyor bu calismalar.. }
%There are other works \citep{Brodie, Constante, Xiao} that aim to detect the inconsistencies in the privacy policies using similar grammatical patterns. However, PPChecker is superior to most of the other related work that uses manually defined grammatical patterns to discover the trustworthiness of privacy policies. Therefore, the process is less time consuming compared to other work.  \color{black} 

%Bu calismaya referans verenleri de incelemek lazim..}

\citep{slavin2016} detect privacy policy violations by creating a privacy-policy-phrase ontology and by mapping the API methods to the phrases in privacy policies. First, API terminology is extracted by finding the phrases in the API documents and having those phrases mapped to the API method names by two investigators. Second, a privacy policy lexicon is built manually by a group of investigators. Finally, an ontology\footnote{An ontology is a formal definition of entities and the relationship between those entities. Here, the entities refer to the phrases obtained from either the API documents or the privacy policies.\color{black}} is constructed again by a group of investigators manually to define the relationships between the terminology extracted from the API documents and the terminology used in the privacy policies. Therefore, using the method names it would be possible to infer the privacy policy terms that should appear in the policy of the application by creating a mapping between the policy terms and the API methods using the ontology. To detect the privacy policy violations, first all words in a given privacy policy is lemmatized\footnote{Lemmatization is similar to stemming, however the dictionary form of the word needs to be produced during lemmatization whereas only the suffixes are removed from the end of the word during stemming. For example, the stem of the word \textit{studies} is \textit{studi}, whereas the lemma form is \textit{study}.  \color{black}}. Then, the lemmatized words and the phrases in the ontology are mapped to find a list of API methods, where the rest of the API methods which do not appear in the privacy policy as a mapping in the ontology implies a violation in the privacy policy. \color{black}

MAPS \citep{MAPS} gives some interesting statistics about the privacy practices of the Android applications in Google Play. According to their analysis, only 50.5\% of apps have privacy policy links on their Google Play Store pages. Moreover, MAPS analyzes the privacy policies by tackling the problem as a classification task. TF-IDF vectors are defined with Boolean values, whether a string exists in the policy statement or not and they are classified by SVC classifier. The classes state whether a policy belongs to a 1stParty or a 3rdParty, or a data type such as Location is specified in the privacy policy or not, or even whether a privacy practice is stated in the privacy policy or not (e.g. ``Our app accesses the location data.''). The classification results \citep{Zimmeck2019} are on a high range of negative F1 scores and the authors suggest an overall improvement upon the state-of-art. 
\color{black}

\subsection{Static/Dynamic Analysis} 

Based on the observation that users cannot understand the purpose of permissions based only on descriptions, a recent study focuses on inferring this information from app's code and behaviours \citep{understandingPurpose:2017}. In the static analysis, two types of features are extracted from the code: app-specific features that include permission related APIs, Intents, Content Providers and text-based features. Text-based features are extracted from identifiers (package, class, method and variable names) in the code. TF-IDF vectors of the word roots in the identifiers that are obtained after pre-processing identifiers are taken as text-based feature vectors. All features are collected from custom code, then given to classifiers for assigning apps to one of the categories of purposes of the following two permission uses: contacts and location. Since supervised learning techniques are employed, an application dataset is constructed and the purpose of these permissions' usage is manually labeled. The results show that text-based features are powerful enough for understanding the purpose of the permission's use, app-specific features are found to supportive. Third party libraries used by apps are taken into account in dynamic analysis. If a sensitive data flow is used by a well-known third party library, its purpose is directly matched with that library. Otherwise, text-based features from the call stack are extracted and used as in the static analysis. It is shown that static and dynamic analysis could cover different scenarios and produce over 90\% accuracy by complementing each other. 

\subsection{Discussion}

\begin{figure}
\centerline{\includegraphics[scale=0.5,trim=0 290 40 320]{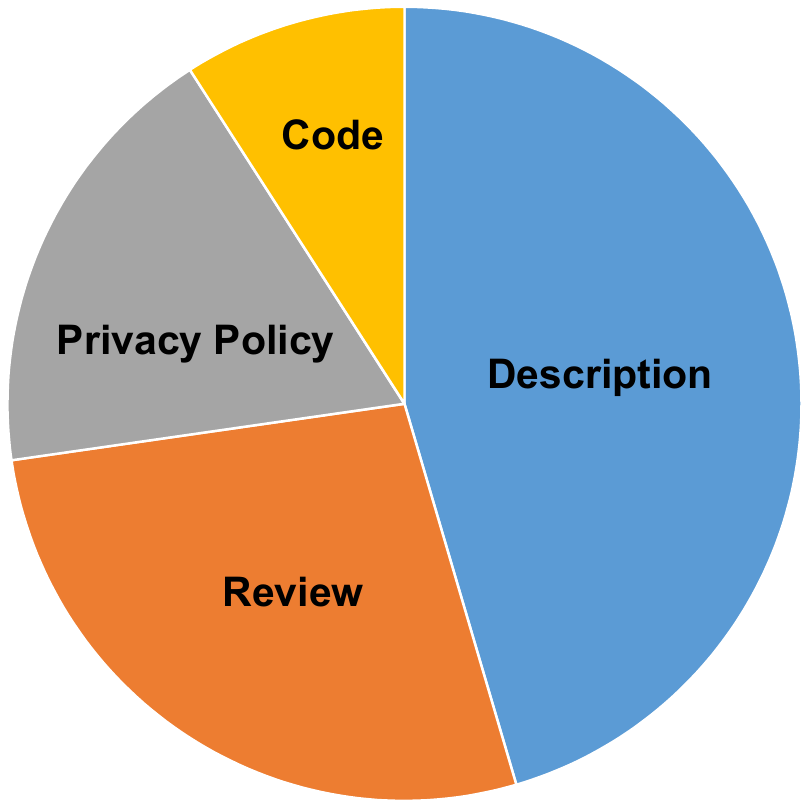}}
\caption{Ratio of Metadata Types in Description-to-Permission Studies}
\label{fig:desctoperm}
\end{figure}

Since 2015, researchers use app descriptions for finding the purposes of dangerous permissions usage in apps and, produce promising results by using NLP techniques. Especially, studies which aim to find semantics through a large corpus (such as word2vec \citep{word2vec}, etc.) and deep learning techniques produce the best results \citep{AC-Net:2019, alecakir2021attention}. Even though most of these studies only use descriptions for extracting the requirement of a permission as shown in Figure \ref{fig:desctoperm}, enhancing these approaches with the usage of privacy policies and user reviews produce promising results in recent studies. Since privacy policies present information about the usage of sensitive data directly in a simple format, we could see more applications of extracting information from privacy policies by using NLP techniques in the future. While the proposed approaches analyze privacy policies on market stores, in-app privacy policies could also be considered \citep{enhancingdtob:2017}. User reviews are also shown to provide useful information about permissions \citep{SmartPi:2019, alecakir2020discovering}. Besides unsupervised techniques, the real effects of user reviews could be observed by introducing labeled datasets that include user reviews. Third party libraries are other resources that could be explored for the usage of dangerous permissions in future studies. The positive effects of using these libraries on decreasing false positives are underlined in \citep{chabadawithTLD:2018}. A recent analysis on third party libraries also claim that the gap between app descriptions and requested permissions is resulted from the extensive use of such libraries in apps \citep{understandingTLD:2017}. Moreover, it is shown that some apps could use more than 20 libraries \citep{privacyGrade} and more than 60\% of subpackages in the code are from third-party libraries \citep{TLDInfo:2015}.  To sum up, other metadata such as privacy policies, user reviews, third-party library documents needs further investigation for the problem in the future. The effects of the quality of such resources on the performance of proposed approaches is open to research. Researchers could evaluate and compare their proposals by using recent public datasets such as AC-NET \citep{AC-Net:2019} and DesRe \citep{DesRe}.  Last but not least, the correlation between such resources and the code could be explored through semantics (e.g.  \citep{localizingFE:2018}).

Besides metadata, app code and behaviours could be analyzed for the problem \citep{understandingPurpose:2017}. A recent study shows that text-based features extracted from the code or the call stack helps finding the reasons of permissions usage in an app. Furthermore, such studies could be used for not only pinpointing the usage of a permission but also proving the requirement of a permission usage declared in app description or in other metadata. This area needs more investigation. Static/dynamic analysis both on custom code and third-party libraries should be considered. While the purpose of a permission usage could be too coarse-grained or fined-grained depending on the technique (static and dynamic analysis respectively) used in \citep{understandingPurpose:2017}, this could be further investigated in future studies.   

\section{Description Generation}

This section covers studies that are proposed to automatically generate security related texts for users such as descriptions, privacy policies, permission explanations, malicious behaviour by using Natural Language Generation (NLG) techniques. NLG is the task of converting any structured or unstructured data into a natural language text. %The task serves to various NLP problems such as simplification of a given complex text \citep{summarising-news}, text summarization \citep{summarization}, automatic generation of questions \citep{question-generation}, automatic generation of reviews for scientific papers \citep{review-generation}, and so on. 
Although template-based methods have been excessively used for the problem \citep{Reiter, McRoy}, statistical methods \citep{langkilde-2000-forest, bangalore, cahill} and recently neural network based encoder-decoder models \citep{Qian, Serban, shen}, autoencoders \citep{CaoClark} have also been used in NLG.

%\textcolor{blue}{Bir de bu calismalar arasinde kendini Drebin ya da baska calisma ile karsilastiran var mi ?}

DescribeMe \citep{DescribeMe:2015} aims to generate descriptions using NLG techniques by explaining any security or privacy related behaviours including permission requests. First, the security behaviour graphs are constructed from a static code analysis. Second, the graph size is reduced with subgraph mining by discovering frequent subgraphs that are peculiar to some specific security behaviours. Eventually, those graphs are converted to natural language text by traversing the vertices in the graphs and converting each path into a sentence based on manually defined English grammar rules under Extended Backus-Naur form (EBNF). Although the results are promising, the number of false negatives is very high as a result of the static code analysis. As the authors note, this could be improved by plugging into more advanced analysis tools. From the description-to-permission fidelity perspective, DescribeMe seems to fail to describe some permission requests because some permissions cannot be resolved, few permissions are not associated with the API calls, and only some permissions are correlated to certain API parameters.  \color{black}

\citep{textMining:2016} applies text mining on human-authored reports by malware analysts and produce explanations of unwanted behaviours in malware using those reports. Therefore malware behaviours are explained in natural language. In the approach, L-1 Regularized Logistic Regression \citep{Tibshirani} is employed as a linear classifier for malware detection by using permissions, actions and API calls as input features. Then, a feature set is defined by obtaining the negative weights from the classifier that indicates an unexpected behaviour, whereas a positive weight indicates a normal behaviour. Based on Beam Search algorithm, the authors design an algorithm to find a subset of the keywords that covers malware instances by ranking them with their TF-IDF weights. Final list of keywords are used to find sentences that include the keywords in malware descriptions and they are ranked by using the cosine similarity between the TF-IDF vectors of the sentences. Automatically extracted description sentences are promising, however those sentences are automatically extracted from the human-authored text rather than being automatically generated as in DescribeMe \citep{DescribeMe:2015}. Moreover, the results of the malware behaviour classification show that DescribeMe can select the features more accurately compared to Drebin \citep{Drebin:2014}.\color{black}

Similar to DescribeMe \citep{DescribeMe:2015}, PRESCRIPTION \citep{personalizedDesc:2019} aims to generate app descriptions from security and privacy perspectives. However, unlike DescribeMe, PRESCRIPTION aims to generate personalized descriptions for different user types. For this purpose, the model learns the user's concerns through their mobile permission changes, where a denied list of permissions are regarded as user's concerns. Then, the users are categorized based on their personalities according to the Big Five Personality model \citep{personality}. For the classification, a questionnaire is published online and a correlation is discovered between the application categories and user types. Moreover, the linguistic preferences are also investigated from different user categories by analyzing the word types used frequently by each user category (such as positive words are used more often by extravert users). Finally, the text is generated based on the user's concerns and the linguistic preferences. During generation, Deep Syntactic Structures (DSyntS) \citep{deepsynts} are used to generate syntactic patterns, where the user-specific parameters help to choose the right pattern for a given user category. The model gives promising results from various perspectives, such as semantics correctness, readability, personality judgement, and system efficiency. Moreover, the generated descriptions are better than that of Drebin \citep{Drebin:2014} in terms of readibility. \color{black}

CLAP (Collaborative App Permission Recommendation) \citep{permReqRec:2018} is a recommender system that recommends a list of potential requirements for the permission explanation. The framework first finds the similar apps to a given app by using an interpolation between the similarities of the descriptions, titles, permissions, and categories. In order to assess the similarities between descriptions, once the stop words and background sentences are removed, the remaining text is stemmed, and unigrams/bigrams are extracted. A text retrieval method called Okapi BM25 \citep{okapi} is used to learn the scores between the description of the current app and the other apps' descriptions in the dataset. Word2vec \citep{Mikolov2013} is used to estimate the semantic similarity between the titles, the Jaccard distance is used to assess the similarity between the permissions, and the cosine similarity between the TF-IDF weighted vectors of the categories is used to estimate the category similarity. In order to identify the permission-explaining sentences in a description, Stanford PCFG parser \citep{stanfordparser2} is used to break the sentences at the conjunctions and verb phrases are extracted to find a set of candidate sentences. The sentences that address the given permissions are found by pre-defined rules based on PoS tags and keywords. Once the explaining sentences are selected, the recommending sentences are chosen by using majority-voting principle on the descriptions that are summarized using a text summarization algorithm called TextRank \citep{textrank}. The results show that CLAP can recommend relevant sentences for permission usages. \color{black}

Unlike the other work, A\texttt{uto}\texttt{PPG} \citep{AutoPPG} generates privacy policies for Android apps by using static code analysis and NLP techniques. It aims to generate easy-to-understand texts that inform users about how their information will be collected, used and disclosed. Fist, API documents are analyzed in order to identify personal information used by APIs. ESA \citep{ESA} is used to compute the semantic relations in the text. Then, by employing static code analysis, invoked sensitive APIs and their mappings to personal information are extracted. The information about under which conditions these APIs are invoked and how the personal information is processed (e.g. stored, transferred) are also obtained in this step. Finally, NLP techniques such as generating sentences based on a template, removing sentences with the same meanings, rearranging sentences in a paragraph according to their importance are employed to generate privacy policies.

\subsection{Discussion}

The studies under NLG with the security or privacy concerns could be categorized into description generation, privacy policy generation, and malware behaviour explanations. The main concern in the description generation research is the lack of the security related information in the description data. Studies such as DescribeMe \citep{DescribeMe:2015} fills this gap by generating security-centric descriptions that will help the app user to be more aware of any security issues of an app, such as the permission requests. A recent study PRESCRIPTION \citep{personalizedDesc:2019} advances this even further by generating personalized descriptions for each user type. A natural outcome of this research is to bridge the gap between the descriptions and permissions. Studies such as \citep{permReqRec:2018} also address to bridge the gap between the descriptions and permissions, but this time by performing as a recommender system that describes the reasons of the usage of a permission, again by using the descriptions as a reference. Another gap in generated descriptions is resulted from the lack of information that needs to collected about third-party libraries. These studies should be extended with third-party library documents and their API documentation. 

A similar concern also lies in the studies that aim to generate malware behaviour explanations. The current studies focus only on malware detection without providing any reasons for the detection result. However, it would be more useful to have an explanation about why an app is considered as malware. This would help to decrease or reverse the damage caused by the malicious application. It would also help malware analysts by reducing their analysis time. \citep{textMining:2016} address this problem by automatically generating malware behaviour explanations. Recently privacy policies have also been generated automatically to inform the user about the information that he/she provides to the app \citep{AutoPPG} with the privacy concerns. % unlike the other studies that have security concerns. 

We expect to see more work on automatic generation of privacy policies. The enforcement by authorities such as Google's announcement on privacy policies \citep{GooglePrivacy:2017} is expected to lead developers to create useful and understandable privacy policies. Since privacy policies are suggested to use a simple language, such policies could be simply generated by template-based methods \citep{Reiter, McRoy}. Such texts could be generated by both using information given by developers or information obtained as a result of the code analysis. Many useful information such as the requirement of dangerous permissions, the flow of private data could be included in policies, hence contribute to solve the description-to fidelity and privacy problems together.

\color{black}

\section{Privacy}

One of the main threats in Android apps is the leakage of private data such as contact lists and user location. In the literature, conventional taint analysis \citep{taintdroid:2014, flowdroid:2014} is generally proposed in order to provide users' privacy. However here, new approaches based on NLP are investigated.  

SUPOR \citep{supor:2015} complements the privacy studies in the literature by identifying sensitive user inputs. It has three major steps: layout analysis, user input (UI) sensitiveness analysis and variable binding. In the layout analysis, input fields of apps are listed. In the UI sensitiveness analysis, a keyword-based approach is applied to input text labels that are closest to input fields for finding sensitive inputs. This is the only step that adapts NLP techniques. Sensitive keyword dataset is constructed by using text in the resource files collected from 54,371 apps. First, each text line is transferred into a parse tree by performing syntactic parsing  with Stanford parser \citep{StanfordNLP:2014}, then nouns and noun phrases are identified by level-based traversals on the parse trees. Finally, sensitive keywords are determined manually. However, during determining the keywords, the dataset is expanded by adding the synonyms of the keywords from WordNet \citep{Wordnet}, and moreover by translating the keywords into Chinese and Korean languages by using Google Translate. Although this lightweight keyword-based approach is promising for sensitive inputs, context-sensitive semantic analysis could be employed for further improvements. The last step associates sensitive user inputs with variables in the code. 

UIPicker \citep{UIPicker:2017} aims to identify user input privacies that are related to account credentials, user profiles, location, and financial activities. UI screens are analyzed to identify any sensitive input. First, pre-processing is applied on the textual data obtained from the user inputs and layout descriptions using NLP tasks such as word segmentation, stemming, and redundant content removal. Chi-Square test is applied to cluster the user input terms that are either privacy related or not. Then, the extracted privacy-related terms on user input and layout descriptions are used as features in Support Vector Machine (SVM) to detect any privacy-related data. In the final step, irrelevant items detected by the classifier in the results are further filtered out by analyzing its behaviour in its program code. The results show that UIPicker can efficiently detect user input privacy data, although it does not consider dynamically generated user input items. 

ClueFinder \citep{findingClues:2018} applies NLP to identify sensitive data on program elements. Once a list of privacy-related keywords is collected from the Google Privacy Policies \citep{google-privacy} and other privacy-related research \citep{financial-times, demetriou, UIPicker:2017}, it is extended by using word2vec \citep{word2vec} and Google Play apps. Then, these keywords are matched with program elements such as variables or methods inside the code, which are the initial sensitive tokens. To further filter out the sensitive tokens, stemming, PoS tagging, and dependency parsing are applied and thus program elements that include sensitive tokens but not related to privacy content are filtered out. %Stemming helps reducing the sparsity in the data by mapping several forms of the same word to a single form, PoS tagging is used to find a specific syntactic usage of a word, and dependency parsing is used to  decide whether a privacy-related keyword is indicative in a sentence. by applying those syntactic tasks. 
Then, all method invocations are analyzed and features such as method name, parameters, and return type are used as features to be used in the classifier to detect any privacy data. Finally, a data flow is performed to check whether private data is passed through unauthorized third parties. As for the evaluation, high precision scores are obtained. BIDText \citep{bidtext:2016} also searches for sensitive keywords inside the code. However, it does not utilize any syntactic task. The results show that ClueFinder has a higher coverage in sensitive source discovery compared to BIDText. 

%Unlike the other work that aims to detect sensitive user input, 
Pluto \citep{Pluto:2016} explores whether the app exposes any user data to  advertising libraries. Such data exposure attacks are handled in two attack channel categories: in-app and out-app. In-app attacks are dependent on the ad library's host app and out-app attacks are independent of the host app. In this study, only the detection of in-app attacks which includes NLP-based techniques are reviewed. The framework flows in two steps: First, an app is run on an emulator by simulating user inputs by using monkey \citep{monkey}, thereby extracting the runtime generated files. %NLP stage applied on manifest files, SQlite database files, runtime generated XML files, and for resource and layout files.
Second, data points are extracted from those files and matched with their synonyms by using Wordnet \citep{Wordnet}. However, a data point can be mapped to several different synonyms depending on the context. For example, the word \textit{exercise} might correspond to \textit{workout} if it is a fitness app, however it might be used in an education app as well, and in that context its snynonym would be different. In order to disambiguate a word, domain knowledge is extracted from the file name or from the table names. If the similarity between a word and its domain knowledge is above a threshold value, then the matching is accepted. Hence, the data points are disambiguated\footnote{The task is called Word Sense Disambiguation, where a particular sense of a word is identified in a given context.\color{black}} within that context.  A similarity metric derived from the LESK \citep{Lesk} algorithm is used to find the similarity between the terms based on the Wordnet dictionary. 

GUILeak \citep{guileak:2018} aims to determine any privacy leakage from user input data and to detect whether any leakage violates the privacy policy of a given app. The framework uses an ontology which is built manually by first creating a phrase dictionary extracted from a set of privacy policies and then defining the semantic relationships between those phrases. A program analysis toolkit called GATOR \citep{Gator} is utilized to extract user input views from the application's code along with a data flow analysis. Those user input views are labelled using a string analysis technique \citep{string-analysis} on the arguments of the API method invocations and dialogs are inserted into their parent activities so that dialog titles and user input labels could be identified for each parent. Policy phrases and input fields are hierarchically mapped to the ontology concepts by using Wordnet \citep{Wordnet} and the cosine similarity between the word vectors of the phrases and  the input fields. Eventually, any policy violation is detected using the input data flow and the results show that privacy policy violations through the user input data can be detected effectively by the framework.

%\color{green}Burcu:Invetter'i yazdim ama daha cok security icerikli gibi geldi bana. \color{red}

%Unlike the other work related to privacy, 
Invetter \citep{invetter} mainly focuses on security-focused input validations. Not all the user input validations involve necessarily security checks. For example, a user input could be just compared to a predefined input format to ensure it is in the correct form. 
Invetter combines static analysis with machine learning methods to detect insecure input validations. NLP is exploited to preprocess the input parameters through word segmentation, stemming, and variable name normalization so that unstructured, ill-defined and fragmented textual data is cleaned and becomes easier to process. Once these preprocessing steps are performed, a total number of 1132 input validations are obtained. In order to detect the input validations that involve any sensitive input, the nearby sensitive input validations are observed since developers normally locate similar input validations in adjacent places. Therefore, a small seed of sensitive input validations are built by investigating whether an input validation verifies the user identity or restricts the usage of sensitive resources in order to discover other sensitive input validations using association rule mining. The results show that Invetter has a decent coverage in identifying the sensitive input validations. \color{black}

Differently from the other related work that addresses the privacy leakage problem that occurs through the user input, \citep{zimmeck2016} analyze the privacy policies to detect any inconsistency between the policies and the actual actions of the app. To this end, the authors classify the privacy policies by using the TF-IDF vectors that are built by the keywords and the bigrams of the keywords in the privacy policy statements. For example, for the location sharing practice, the authors use keywords such as  \textit{geo} and \textit{gps} to find all the relevant sentences in the privacy policies, and the actions in those sentences are gathered along with their bigrams. Eventually SVM and logistic regression are used for the classification. Those analysis results are compared to what actually an app is doing through a static analysis. Their evaluation shows that the proposed approach is reliable and the accuracy is above the baseline.  

%Libert \citep{Libert} -> nlp yok.

\color{black}

%\textcolor{red}{\textcolor{red}{Pluto \citep{Pluto:2016} Bunu citep edenlere de bakmak gerekir baska ilgili calisma var mi diye.. }\\
%\textcolor{red}{Invetter: Locating insecure input validations in Android services su calisma alakali bakmak lazim, cok iyi yerde yayinlanmis.. }

%\textcolor{blue}{BAKILMAYAN YORUM: GUILeak \citep{guileak:2018} : ui ile privacy'i iliskilendiriyor. Referans verdigi ve karsilastirdigi calismalar da var, ama bu calismalar ui ile ilgili degil api ile privacy eslestirmesine bakiyor : \citep{slavin2016}, \citep{zimmeck2016}. Yani bu calismalari baska baslik altinda bakmak gerekir mi bakmamiz lazim. Policyler ile ilgili NLP kullanilan baska calismalar da var: \citep{policylint:2019}}

\subsection{Discussion}

In this section, the studies that aim to prevent any leakage of private data through the user inputs are revised under the privacy concerns. Although the overall detection results are promising in this category of studies, there are still limitations of the frameworks. For example, if the data is encrypted, then the privacy violations may not be detected \citep{guileak:2018}, or if the user provided data is stored in a file, again the violation may not be detected \citep{guileak:2018}. Moreover, keyword-based approaches without trying to find the relations between the words/phrases within a sentence may not extract relevant privacy-related keywords, which may lead to false positives. Therefore, most of the privacy-related studies use dependency parsing to extract the privacy-related keywords \citep{findingClues:2018, supor:2015, asdroid:2014}. Those keywords are expanded by using either Wordnet \citep{asdroid:2014, supor:2015} or word2vec \citep{findingClues:2018} for the synonyms. It is also not possible to detect the sensitive keywords without understanding the Android framework. Therefore, it is a good idea to combine the inferred semantics using the potentially sensitive keywords with other information extracted from the Android framework, such as the package name. \color{black}

In the future, more recent deep neural network architectures could be applied for extracting the keywords and the relationships between them. For example, attention networks \citep{Sutskever, self-attention} have shown a significant improvement on almost any NLP task (e.g. machine translation) and they could be utilized for the privacy-related problems. Those networks enable to find the important bits of information in any task. For example, in machine translation, during the translation of a word in a target language, the attention (i.e. the contribution) will be much higher for the equivalent word in the source language. Therefore, a similar approach also could be applied to spot the sensitive keywords especially for the privacy problem. \color{black}

A further step in the privacy-related research would be also the automation of the ontology construction, which will save considerable amount of time. % to the researchers in the field.  

\color{black}

\section{Malware Detection} 
\label{sec:md}

In recent years, NLP techniques are employed to enhance malware analysis and detection techniques in Android. Such proposals are studied under three groups here. The first group covers proposals for assessing description-to-behaviour fidelity in order to detect outliers in different app categories or to find suspicious apps which need further analysis to tag them as malicious or benign. Some of these studies are also covered under description-to-behaviour fidelity proposals above, however here, their ability on malware detection is put under investigation. Recently, text-based features are taken into account in mobile malware detection, the second part studies how these proposals employ NLP-based techniques in order to extract text-based features from app code and/or app metadata. In the last  group, studies that focus on extracting the profiles of benign and malicious apps by using NLP are brought together. 

\subsection{Description-to-behaviour Fidelity}
CHABADA \citep{Chabada:2014} is a system that checks application behaviour against application descriptions by using static analysis and NLP techniques. Even though it is not claimed to be a malware detection system, it could complement traditional detection techniques. %Therefore, it is analyzed under this section. 
CHABADA firstly defines the topics of collected apps by using  the topic modelling method LDA (Latent Dirichlet Allocation) \citep{LDA}. Then, apps are clustered based on their topics using K-means algorithm. Then, the usage of sensitive APIs for each cluster is extracted statically. Finally, the outliers of each cluster with respect to their API usage are detected by using one-class SVM. The main aim  is to find the applications whose behaviour is not consistent with the topic clusters that it belongs to. For example, while accessing your location is normal for a weather app, it might be abnormal for a calendar app. Besides detecting such anomalies, CHABADA is assessed for detecting malwares. The results show that CHABADA could detect 56\% of the malicious apps, obtained from a part of Malgenome dataset \citep{Malgenome:2012}. A recent study \citep{chabadawithTLD:2018} takes into account third-party libraries in order to improve CHABADA \citep{Chabada:2014}. It follows a similar approach as CHABADA, but removes third-party libraries from the code. Hence, it could decrease false positives by filtering out outliers resulting from the third-party libraries. Moreover, malicious code %in the custom code
could be pinpointed.

%A recent study claims that the gap between app description and requested permissions is due to the usage of third-party libraries \citep{understandingTLD:2017}. Such libraries could be responsible for most of the permission uses in the code. Moreover, it is shown that some applications could use more than 20 libraries \citep{privacyGrade} and more than 60\% of subpackages in the code are from third-party libraries \citep{TLDInfo:2015}. Therefore, a recent study \citep{chabadawithTLD:2018} takes into account third-party libraries in order to improve the results of CHABADA\citep{Chabada:2014}. It follows a similar approach as CHABADA, but removes third-party libraries from the code. Hence, it could descrease false positives by filtering out outliers resulting from third-party libraries. Moreover, malicious code in the custom code could be pinpointed. 

%\textcolor{blue}{\citep{chabadawithTLD:2018} bunun isaretledigim referanslarina bak, BAKTIM}

TAPVerifier \citep{enhancingdtob:2017}, which is mentioned above as an enhancing approach for description-to-behaviour fidelity by using privacy policy of applications, is also explored on malware detection and compared with CHABADA \citep{Chabada:2014}. The experiments are performed on 12 apps out of the malicious apps detected by CHABADA, since only those applications' privacy policies could be obtained. The experimental results show that most of the apps tagged as malicious by CHABADA are false positives and some of these false positives could be eliminated by taking app's description and privacy policy into consideration. Even though, such approaches for assessing description-to-behaviour fidelity could improve malware detection, they need further scrutiny in order to prove that.

\subsection{Using Text-Based Features}

Besides static and dynamic features of apps, metadata (\%3) has been started to be used for malware detection \citep{featureSurvey:2015}. Some recent studies \citep{adroit:2016, metaMalwareDet:2016, malware:2016} use metadata of apps in addition to other static features such as permissions, API calls. In \citep{metaMalwareDet:2016}, category information and description of apps are given as inputs to the SVM algorithm besides permissions and API calls. Descriptions are pre-processed and their TF-IDF representations are generated. By employing LDA on the TF-IDF representations and employing K-means clustering on LDA, the binary representation of the clustering is produced. The classifiers are generated by using different sets of features. When metadata features are provided, both models based on permissions and API calls have improved. While the effect of metadata features is more influential on the permissions-based model, it slightly affects the API calls-based model. ADROIT \citep{adroit:2016} uses app descriptions besides permissions and other metadata collected from the app store (user ratings, number of votes, minimum Open GL and SDK versions) as features for discriminating malware from benign apps. In order to represent descriptions, a corpus for building a term-document matrix is constructed. In order to do that, descriptions are pre-processed by removing special characters and stopwords, stemming and stripping white spaces. In the results, the positive effects of metadata of apps on malware detection are shown on the dataset collected by \citep{Aptoide}.  

\citep{malware:2016} %also investigate the effects of using metadata on malware detection. Applications are 
analyze apps based on two approaches: Semantic Pattern Transformation and statistical-based analysis. They use metadata features  %metadata consists of many attributes of applications %that can be collected from Google Play 
such as description, price, number of downloads, information about developers. The main idea of Semantic Pattern Transformation is to convert value-centric vectors into semantic-aware presentations. This transformation allows to analyze semantic relations between features, to search such semantic relations and to extract relevant features. In this transformation, some basic NLP techniques are applied in order to process descriptions such as removing stopwords, lemmatization, and PoS tagging. In the statistical-based analysis, apps are analyzed based on statistics of metadata such the frequency of words in the applications belonging to the same category. Both analyses support that knowledge extracted from metadata, which could be useful for malware detection and could complement more advanced techniques. Even though the analysis is not integrated with a malware detection system which is left as a future work, it is proposed as an alternative method that does not require applications' packages and execution of applications.  

Another use of app descriptions is proposed for app clone detection \citep{FUIDroid:2017}, since repackaging, where the malicious code is injected into benign apps, is quite popular among Android malware \citep{Malgenome:2012, repackaging}. 
%which is one of the most used techniques among Android malware \citep{Malgenome:2012}. 
In the proposed approach called FUIDroid, at first descriptions are employed in order to find apps with similar functionality. Then similar ones are further investigated for UI similarity. Here, only the first component which includes NLP-based techniques will be explained. In that component, firstly the most used 15,000 keywords in descriptions are extracted by using basic NLP techniques to construct a standard dictionary. Then descriptions are represented as dynamic vectors filled with the index of description keywords in the dictionary. Since distance-based similarity calculation such as cosine similarity is not suitable due to complex computing, a tree-based multi-keyword approach is applied \citep{multikeywordRank:2015}. The experimental results show that  the size of suspicious cloned apps forwarded for UI analysis is importantly reduced by using this method, hence the approach gains scalability.

For supporting such studies, a new malware dataset called RmvDroid \citep{rmvdroid:2019} is recently introduced. RmvDroid not only contains application, but also application metadata such as user rating, category, app desciption, developer name, app installs. 

A study on grayware is given in \citep{grayware:2016}. While malware deliberately harms mobile users/devices, grayware falls in an gray area. It is defined as apps cannot be labeled as malware, but still negatively affects users in terms of privacy, performance and user efficiency and analyzed under 9 categories \citep{grayware:2016}. Two of them are related to installation tactics: impostors and misrepresentators. Impostors are apps that impersonate other apps, especially popular ones for ensuring to be installed \citep{grayware:2016}. A recent study shows that a considerable amount of impostors apps are malicious and among top apps in Google Play \citep{neuralEmbeddings:2019}. Misrepresentators are apps that claim to provide functionality to users that is not implemented in the app code \citep{grayware:2016}. These categories are of particular interest to this study, as they may require NLP techniques to be detected. The same hypothesis is proposed in \citep{grayware:2016} that text analytic techniques are needed besides program analysis for triage such apps. Even the simple text analytic techniques are shown to be useful for decreasing the number of grayware apps to be analyzed by humans \citep{grayware:2016}. Future work could explore more advanced techniques. 

A different approach by enhancing Android malware detection with tweets is introduced in \citep{twitterEnhanced:2017}. %It is the first initiative to use social media for Android malware detection. 
In the study, how to link apps with related tweets are explored at first. TF-IDF and Weighted Matrix Factorization approaches are employed to represent descriptions and tweets as vectors. Then, the similarity scores between documents are computed by using the cosine similarity. The performance of the linking approaches are evaluated on a ground truth dataset, in which apps are obtained from AndroZoo \citep{androzoo:2016} and PlayDrone \citep{playdrone:2014}, and their related tweets collected by using the HTTP-based linking (the existence of a link to the app in the tweet). For malware detection, in addition to traditional features \citep{roy2015experimental}, tweet-based features, namely TF-IDF vector, WMF vector, the sentiment of tweets identified by the Stanford NLP library \citep{StanfordNLP:2014} and metadata of tweets such as the number of followers of the corresponding tweet's author are employed. The results show that if tweets are successfully linked to apps, tweet-based features, especially the ones related to metadata of tweets, are beneficial for improving malware detection.

A recent study \citep{networkFlows:2017} treats HTTP flows of apps as textual documents, inspired by %studies in the literature that observe 
similarities between application protocols and natural languages \citep{semanticProtocol2016}. Since HTTP/HTTPS are the major protocols used by malware, they extract text-level features from HTTP flows in order to classify apps into benign or malware by using SVM. Firstly, HTTP flow headers are divided into words by applying word segmentation. Here, special characters included in HTTP headers are used for segmentation. Then, after cleaning headers from meaningless and stop words, they are represented as n-grams. n is experimentally chosen as 1, which means different words in headers are assumed to be independent from each other. Finally, feature vectors are constructed by representing each flow header as a one-hot vector by only considering presence of words in the header. The experiments produce high detection rate (99.15\%) with a very low false positive rate (0.45\%).% by using these feature vectors. 

One of the earliest studies in Android security that correlates texts in UI with the code is AsDroid (Anti-Stealth Droid) \citep{asdroid:2014}, which defines stealth behaviour as the behaviour that do not match with UI. For instance, a  ``login'' button that leads to a phone call is considered as a stealthy behaviour. The proposed approach has two main steps: static and UI analysis. Static analysis finds a top level function of UI (e.g. onClick() function of a button) that is associated with the behaviour of interest. Six types of behaviours are considered in the paper: sending sms, making a phone call, http access, install, sms notification and ui operation. 
In UI analysis, the intent of the behaviour is matched with UI by using text analysis techniques. Firstly, the text that corresponds to a top level function is extracted from the static layout construction. Then, keywords are parsed by using the Stanford Parser \citep{StanfordParser} in order to create a set of keywords for each intent. Here, keywords or keyword pairs that have the highest frequency and that cover all the top level functions with the intent type of the interest are obtained. The set is enriched by adding synonyms of the keywords by using Chinese WordNet \citep{chineseWordnet}. The final set is checked by a human analyst. Finally, for a top level function with UI text T and the intent I, the algorithm searches T in the keyword set for I. If a match is not found, the behaviour is labelled as malicious. AsDroid produces promising results, hence the research community has moved in the direction of discovering useful information from UI texts for privacy as shown above. While the approach is simple and effective, more advanced text analysis and image analysis techniques for more advanced malware are left as future work. 

\subsection{Extracting Applications Profiles}
An interesting approach proposes a framework called apk2vec in order to automatically build behaviour profiles of given Android apps that could be used for many tasks from malware detection to app recommendation \citep{apk2vec:2018}. In this work, once the static analysis is performed by generating three different dependency graphs obtained from API dependencies, permission dependencies and source-risk dependencies, those dependency graphs are used to train the embeddings using the skipgram architecture of word2vec \citep{word2vec}. Word2vec \citep{word2vec} enforces the words co-occurring in similar contexts to have similar embeddings while training for word embeddings. Therefore, given an input word to the one layer neural network, the output is the predicted context of this word. Analogously, in apk2vec, the apk id is given as input to the neural network and the network is expected to predict the API, permission and source-sink subgraphs that occur in the apk's context. Therefore, the embeddings of the API, permission and source-sink subgraphs tend to be closer to each other if they have the same apk class label, such as malware.  

apk2vec is shown to be useful for many security related tasks. In batch malware detection, SVM is employed to classify the profiles of collected apps from Virus Share \citep{VirusShare}, Drebin \citep{Drebin:2014} and \citep{GooglePlay}. In online learning, apps are firstly ordered according to their release data, then the first part is used to train an online Passive Aggressive classifier. For malware family classification, K-means algorithm is employed for clustering the Drebin dataset \citep{Drebin:2014} into 179 families. All results of these tasks support that apk2vec is task-agnostic and represents app semantically well. Another usage of apk2vec for security is clone detection. A recent study \citep{neuralEmbeddings:2019} employs icon and text embeddings of apps and their similarities to the embeddings of popular apps in order to find impostor apps (counterfeit apps as given in the study). Even though the main contrubiton of the study is to propose a new approach on combining content and style embedding from pretrained convolutional neural networks (CNNs) in order to calculate the similarity of app icons, it is shown that the results are increased by 3\%-5\% (precision) and 6\%-7\% (recall) by including text embeddings of app descriptions.

\begin{table}[t]\vspace{2pt}
\scriptsize
\caption{Outline of Related Studies}
\begin{center}
\begin{tabular}{| c | c | l |}
    \hline 
    \textbf{Problem} & \textbf{Sub-Category} & \textbf{Related Work} \\
    \hline 
    \multirow{6}{*}{DESCRIPTION-to-BEHAVIOUR} & Description & Whyper \citep{Whyper}\\
    & & Autocog \citep{Autocog} \\
    & & ACODE \citep{ACODE:2015} \\
    & & AC-NET \citep{AC-Net:2019}\\
    & & DesRe \citep{alecakir2021attention} \\
    \cline{2-3}
    & Review & AutoReb \citep{autoreb:2015}\\
    & & \citep{shortText:2019}\\ 
    & & SmartPI \citep{SmartPi:2019} \\
    & & LSTM-based \citep{alecakir2020discovering}\\
    \cline{2-3}
    & Privacy Policy & TAPVerifier \citep{enhancingdtob:2017} \\
    & & PPChecker \citep{trustpolicy:2016}\\
    & & \citep{slavin2016}\\
    %& &  \\
    \cline{2-3}
    & Code  &  \citep{understandingPurpose:2017} \\
    %\cline{2-3}
    \hline
    \multirow{3}{*}{DESCRIPTION GENERATION} & Description & DescribeMe \citep{DescribeMe:2015}\\
    & &  PRESCRIPTION \citep{personalizedDesc:2019}\color{black}\\
    \cline{2-3}
    & Privacy Policy & AutoPPG \citep{AutoPPG}\\
    & & MAPS \citep{MAPS}\color{black}\\ 
    \cline{2-3}
    & Permission Explanation & CLAP \citep{permReqRec:2018}\\ 
    \cline{2-3}
    & Malware Behaviour & \citep{textMining:2016} \color{black}\\
    \hline
    PRIVACY & N/A & UIPicker \citep{UIPicker:2017}\\
    & & SUPOR \citep{supor:2015}\\
    & & ClueFinder \citep{findingClues:2018} \\ & & Pluto \citep{Pluto:2016}\\
    & & GUILeak \citep{guileak:2018}\color{black}\\
    & & Invetter \citep{invetter}\\
    & & \citep{zimmeck2016}\\
    \hline
    \multirow{3}{*}{MALWARE DETECTION} & Description-Behaviour &  CHABADA \citep{Chabada:2014}\\
    & & CHABADAwithTLD \citep{chabadawithTLD:2018} \\ 
    & & TAPVerifier \citep{enhancingdtob:2017}\\
    \cline{2-3}
    & Using Text Features & ADROIT \citep{adroit:2016}\\
    & & \citep{metaMalwareDet:2016}\\
    & & \citep{malware:2016} \\
    & & AsDroid \citep{asdroid:2014}\\
    & & FUIDroid \citep{FUIDroid:2017}\\
    & & Grayware \citep{grayware:2016} \\
    &  & Twitter-Enhanced \citep{twitterEnhanced:2017}\\
    & & Network Flows \citep{networkFlows:2017}\\
    \cline{2-3}
    & Extracting App Profiles & apk2vec \citep{apk2vec:2018}\\
    & & FeatureSmith \citep{featuresmith:2016}\\
    %\cline{2-3}
    \hline
\end{tabular}
\end{center}
\label{tab:outline}
\end{table}

FeatureSmith \citep{featuresmith:2016} differs from other NLP-based malware detection systems by proposing an approach for automatic feature engineering. It adopts scientific papers in the literature as the source of information and mines them for extracting malware features. Even though it is a generic approach, Android malware detection is selected as the proof-of-concept study. It extracts 195 features by mining 1068 scientific papers on the area. For the feature extraction, at first, the Stanford dependency parser \citep{stanford-parser} is used to extract the dependencies that correspond to basic malware behaviours. Those dependencies are used to build an ontology where each node corresponds to the behaviours, malware families and concrete features and the edges correspond to the relations between the entities. Edges are also weighted by the semantic similarity between the entities, so that features could be quantified based on its relevance to detect Android malware. \color{black}It is compared with Drebin \citep{Drebin:2014} which employs 545,334 manually engineered features. Random forest classifiers are trained with these feature sets and compared. The results show that FeatureSmith reaches to the similar performance with Drebin, even with very small set of features. Furthermore, it could detect some malware families (e.g. Gappusin) that Drebin cannot detect. 

\subsection{Discussion}

Although the studies reviewed in this section show that using NLP-based techniques and text-based features can help improve malware detection, they are at an early stage. While CHABADA \citep{Chabada:2014} and TAPVerifier \citep{enhancingdtob:2017} have mainly proposed for assessing the description-to-behaviour fidelity, their effects on malware detection are also evaluated, but in a very limited way. Such evaluations, hence such studies can be further improved by using larger datasets. From this point of view, introduction of malware datasets including metadata such as RmvDroid \citep{rmvdroid:2019}, AndroZoo++ \citep{androzoo:2016} are good initiatives for accelerating research on this domain. Furthermore, topic-specific security solutions \citep{Chabada:2014, characterizing:2017}, which employ LDA-based techniques on app descriptions for finding apps' topic, helps characterizing malware based on topics they belong to.

%\textcolor{blue}{BAKILMAYAN YORUM: burada descriptionlarin vektorleri falan cikarilabilir mi Burcu ile tartisalim..more advanced techniques ne olabilir.}
In malware detection, researchers mainly use text-based features extracted from descriptions \citep{adroit:2016, metaMalwareDet:2016, malware:2016}. Other metadata such as user reviews, privacy policies processed by NLP techniques could be employed to enrich feature vectors. Social media could be another source that could be exploited for obtaining text-based features \citep{twitterEnhanced:2017}. Moreover, more advanced NLP techniques could be employed to extract features, detect malware and grayware such as misrepresentators. 

Deep learning techniques with word embeddings have shown dramatic improvements in many NLP tasks. The popular word embedding model word2vec \citep{word2vec} has inspired many new vecs to emerge such as apk2vec \citep{apk2vec:2018}, spam2vec \citep{spam2vec}. The main idea of \textit{something}2vec studies is that similar somethings of the same type (e.g. words, spams) could be represented by vectors located close to each other in the vector space. Similar idea could be used for the purpose of Android security from malware family detection to repackaging detection by representing malware as vectors by using rich information of apps. apk2vec \citep{apk2vec:2018} based on this idea extracts Android app profiles. In the future, such studies could be enriched by using more information such as metadata with the aim of malware detection.  

%\section{Discussion}

%other metadatanin kullanilmasi (Android)

%farkli teknikler (monitoring, tracking), farkli uygulamalar icin kabul edilebilir ya da illegal olabilir..  

%desc-to-behaviour'un malware detection uzerine etkisinin daha cok arastirilmasi. CHABADA, TAPVerifier CHABADA tarafından etiketlenenlerin false positives oldugunu gostermis. .. 

%*** otomatik privacy generator from the code.. 

\section{Related Work}
\label{sec:RW}

In this section, traditional security mechanisms proposed for Android security are presented shortly. Two common types of malware analysis and detection techniques are static and dynamic analysis. Static analysis analyzes the code and the application package without running the code. On the other hand, dynamic analysis monitors the runtime behaviours of apps. Permissions and API calls are among the most used features in static analysis \citep{featureSurvey:2015}. Static analysis is not resilient to some evasion techniques such as obfuscation \citep{droidchameleon, sen2018coevolution}, dynamic code loading \citep{xue2017auditing, aysan2018analysis}. Hence such techniques are not very successful at detecting new attacks or even new variants of existing attacks. There are also effective evasion strategies against dynamic analysis. For example, a malicious app could hide its malicious behaviours when it identifies being run in a virtual machine environment \citep{rage:2014, evadingSandbox:2014}. While a bare-metal analysis could be a solution to this problem \citep{baredroid:2015}, such systems are generally not very efficient for analyzing a large number of apps. Triggering malicious code part in apps is another issue for runtime analysis. Even though there are proposals on automatically triggering malicious code in Android \citep{grodddroid:2015, intellidroid:2016, droidbot:2017}, they might not be very effective for all kinds of malware \citep{inputgen:2015}. In addition, dynamic analysis might not be affordable on some devices due to their resource constraints, especially in terms of power. Therefore, security solutions on mobile devices known as anti-malware systems mainly rely on static analysis.  %It is also the same from the academic point of view, since most of the studies in the literature propose static analysis-based solutions for the problem. 

As it is summarized, both techniques have advantages and disadvantages. There exists also hybrid approaches that combine both techniques \citep{harvester:2016, ec2:2017, updroid:2018}. All these techniques could be applied at both on device or on the app stores. Some markets analyze submitted apps before making them available for download and install. Taint analysis is another effective technique for detecting data leaks in mobile devices through constructing data flows statically or dynamically. Taintdroid \citep{taintdroid:2014} and FlowDroid \citep{flowdroid:2014} are among the well-accepted studies in this area. Please see a recent survey of Android malware and analysis techniques for further information \citep{malwareSurvey:2017}.

One of the earliest work on exploring the usage of NLP for security \citep{atallah2000natural} divides the possible applications of NLP in information assurance and security into four groups:  NLG for memorizing random strings such as strong passwords, watermarking natural language text, applying machine translation for information security (e.g. translating a text into another language before encrypting it), and sanitizing information. However, since this publication there have been many advancements in both information security and NLP. For example, there are recent promising approaches of NLP in cyber security from cyber threat intelligence \citep{cti:2016, chainSmith:2018} to anomaly detection using sequential data such as logs \citep{deeplog:2017} and social media \citep{smAD:2018}. 

The deployment mechanism of mobile apps has recently attracted researchers to explore the application of NLP techniques in this research area. There are already few review studies on application store analysis for software engineering \citep{sesurvey:2016, revsurvey:2017}. However it is believed that there is also a need to review the literature from the security point of view. This is the main aim of the current study.

\section{Discussion}
\label{sec:disc}

The discussion section is covered under three subsections. Firstly, the general discussion of existing NLP-based studies on Android security is given. Then, the strengths and weaknesses of the NLP techniques in Android security are discussed. Finally, possible directions for future research are summarized.

\subsection{NLP-based Approaches in Android Security}
The studies on the use of NLP methods for Android security has been exploring for the last five years, so it is still in its early stage. In this study, we have given a survey of research on NLP-based techniques for Android security for the following four research problems: description-to-behaviour fidelity, description generation, privacy and malware detection, and the proposals are discussed separately for each problem. The studies in the literature are summarized in Table \ref{tab:outline}. In this section, a general view of the proposed studies is aimed to be represented. Moreover,  how they could be employed to complement traditional security solutions will be discussed.

While the studies mainly focus on application descriptions, other type of metadata in textual form, especially privacy policies and user reviews, need to be the subject of further studies. The use of privacy policies is a good candidate from the security point of view. Moreover, the enforcement on the use of privacy policy by Google is expected to make such approaches more applicable in the near future. Furthermore, these documents are expected to be simple, hence simpler to be processed by NLP-based techniques. Appropriate weight should also be also given to user reviews, which directly reflect the experience of users on applications. The effects of different type of metadata on Android security should need to be explored. 

Adversarial attacks against NLP-based systems in Android is an open research area. With the increasing popularity of deep learning methods in NLP, such methods have also become the target of attackers \citep{wang2019survey, zhang2019adversarial}. We believe that with the increase in the
use of NLP techniques in Android security, new attacks against such systems are going to emerge. Evasion and poisoning attacks are among the mostly studied attacks in other domains \citep{corona2013adversarial}. In Android security, how attackers could evade from the NLP-based solutions and how they could poison the training of an NLP system with well-crafted data needs to be studied. For instance, the effects of fake reviews should be analyzed and researchers should address how to select effective user reviews. 

%Text generation is still an immature research area.
There would appear to be insufficient research on text generation. Automatic generation of descriptions or privacy policies could help developers to generate user friendly metadata. Furthermore, if some rule-based approaches are enforced by market stores for this problem, a generic form of metadata could be generated, which helps to process the text. Such techniques could also be used for the explanation of mobile malware and complement works on malware familial classification based on program analysis \citep{updroid:2018, andrensemble:2019}. Such approaches could give an idea about the detected malicious application to malware analysts, hence helps to reduce their time on malware analysis. Furthermore, if the behaviour
of a detected malware is known, specific steps can be taken to decrease or reverse the damage caused by the malware.

We believe that NLP-based techniques could complement the traditional approaches in the literature well. If we find inconsistencies between permissions and user metadata, this would be a strong indication of a malicious application. Then, more resource-intensive analysis techniques could be applied on such apps. Moreover, traditional code analysis techniques could be enhanced with semantic analysis by considering some part of the code as text (e.g. method names, API calls).

\subsection{Strengths and Weaknesses of the NLP Techniques in Android Security}

As reviewed in this article, a wide range of NLP techniques have been used for the problems relevant to Android security. NLP has evolved substantially in the last decade by having most of the statistical-based methods extinct and transforming the traditional methods into deep learning-based models. Other research fields like Android security have their share during this evolution. While early work on Android security was using statistical-based models such as tf-idf to obtain the vectorial representations of words to define their semantics, the recent work is using more neural representation models such as word2vec \citep{word2vec}. However, frequency based models such as tf-idf are still being widely used by the Android security field. 

Learning semantics is one of the key problems under Android security for almost all of the problems such as description-to-permission fidelity, privacy policy, or description generation. Tf-idf is a simple scoring scheme and it only defines how important a word is in a document. It gives some clues about the semantics of a word, but it is not as powerful as the neural representation models. On the other hand, neural representation models have the ability to capture all semantic and also syntactic features of a word and can effectively define everything in a low-dimensional vector, whereas tf-idf is a linear vector where the size of the dimension is equal to the vocabulary size in the corpus. Therefore, it brings also performance issues, which is quite important in tools developed for Android security. Feature extraction and dimension reduction need to be performed to alleviate the performance problem in such cases. From the adversarial attacker point of view, the methods which are directly based on the frequencies or the occurrences of words have the potential to be deceived, whereas neural models can build a compact representation of any text and it could be more stable compared to frequency-based models. However, from the other hand, neural models are highly dependent on the data that they are trained on. Rather than a general corpus, it would be more robust to train such models on domain specific corpora on security related texts. 

WordNet \citep{Wordnet} has also been widely used by researchers on Android security in order to extract semantic information from textual data and it is still used in security. The major limitation of WordNet is that it is restricted with its hierarchical representation and since it is hand-crafted, it is built for general purposes and it is not domain-adapted. However, neural models such as word2vec \citep{word2vec} can be easily trained on a domain specific corpus and more specific neural word embeddings could be obtained. If any hierarchy (e.g. synonym, hypernym, or hyponym) is required along with the meaning, then ontology-based corpora such as WordNet could be more beneficial, since it is harder to obtain this kind of hierarchies in neural word embeddings. However, if solely semantic similarity or relatedness is required, then neural word embeddings are more effective compared to ontology-based models.

Most of the security studies apply preprocessing on the textual data to reduce the sparsity in data. For example, in almost any task, the stopwords are removed since they do not contribute much on the meaning. However, this might have a side effect by substantially changing the overall representation of the text.  For example, the meaning of the description statement, ``Contact your group to the new boss'' is different than the stopword-filtered statement ``Contact group new boss''. 

In addition to the requirements under semantics, it is essential to discuss on the syntactic level as well. In the syntax level, studies usually perform PoS tagging and dependency parsing. Most of the studies usually adopts these tasks from external sources such as Stanford parser \citep{stanfordparser2}. Those external sources are quite successful for especially English language, and it can accurately find the relations and dependencies between the phrases in a given sentence. Since, provided textual data is quite formal in Android security, the common noise problem in NLP is not an issue in the field. We have not come across any study that uses chunking but it would be also helpful to perform verb phrase or noun phrase chunking to find the relevant actions or objects in a sentence, when specific actions or objects are required to be found in a task, which could be another potential syntactic processing to be applied on the textual data provided by Android security.

%The preprocessing (especially stemming) has a significant impact on permission classification task. There are sentences such as S2 which is labelled as a statement sentence. Due to stopword removal process, some of the important tokens, which may lead directly to a decision, are removed. In this example sentence S2, pronoun "your" is removed in stopword removal step, and then the resulting sentence becomes "contact group new boss". Therefore, the resulting sentence has a different meaning than the original. Including syntactic information such as part-of-speech tag would mitigate these types of errors in the model, which remains as future work.
%YORUM: Bir improvement olarak, PoS tag de embedding'lere eklense belki bu tur hatalari cozebilirdi. \color{black}

\begin{table}[h]\vspace{2pt}
\tiny
\caption{Comparison of the NLP methods used in Android security}
\begin{center}
\begin{tabular}{| l | p{2.4cm} | p{2.9cm} | p{4cm} | }
    \hline
    \textbf{Models} & \textbf{Advantages} & \textbf{Disadvantages} &  \textbf{Related Work}\\
    \hline
    \multirow{1}{*}{Rule-based} & Good performance & Requires too much effort & PPChecker \citep{trustpolicy:2016}\\
    & & to define the rules, & DescribeMe \citep{DescribeMe:2015} \\
    & & low coverage & \\
    \hline
    \multirow{1}{*}{Statistical} & Do not require  & Fragile to adversarial & ACODE \citep{ACODE:2015}\\
    & large amount of data &  attacks, comparably  & AutoReb \citep{autoreb:2015}\\ & & lower accuracies & SmartPI \citep{SmartPi:2019}\\ & & & \citep{shortText:2019}\\ & & & MAPS \citep{MAPS} \\ & & & \citep{understandingPurpose:2017} \\ & & & \citep{textMining:2016}\\ & & & PRESCRIPTION \citep{personalizedDesc:2019}\\ & & & CLAP \citep{permReqRec:2018}\\ & & &  UIPicker \citep{UIPicker:2017}\\ & & & ClueFinder \citep{findingClues:2018}\\ & & & Invetter \citep{invetter}\\& & & \citep{zimmeck2016}\\ & & & CHABADA \citep{Chabada:2014}\\ & & &  \citep{metaMalwareDet:2016}\\ & & & ADROIT \citep{adroit:2016}\\ & & & \citep{malware:2016}\\ & & & FUIDroid \citep{FUIDroid:2017}\\ & & & \citep{twitterEnhanced:2017}\\ & & & \citep{networkFlows:2017}\\
    \hline
    \multirow{1}{*}{Corpus-based} & More robust in   & Requires an annotated  & Whyper \citep{Whyper}\\
    & semantic meaning & dictionary or corpus & AUTOCOG \citep{Autocog}\\
    & extraction & (such as WordNet & SmartPI \citep{SmartPi:2019}\\
    & & or another ontology) & TAPVerifier \citep{enhancingdtob:2017}\\
    & & & \citep{slavin2016}\\
    & & & AutoPPG \citep{AutoPPG}\\
    & & & SUPOR \citep{supor:2015}\\
    & & & Pluto \citep{Pluto:2016}\\
    & & & GUILeak \citep{guileak:2018}\\
    & & & ASDroid\citep{asdroid:2014}\\ 
    & & & FeatureSmith \citep{featuresmith:2016}  \\
    \hline
    \multirow{1}{*}{Neural} & Easier to implement & Prone to fail for & AC-Net \citep{AC-Net:2019}\\
    & High accuracies & the noisy data, requires & DesRe \citep{alecakir2021attention}, LSTM-based \citep{alecakir2020discovering}\\
    & & large amount of data & apk2vec \citep{apk2vec:2018} \\
    \hline
\end{tabular}
\end{center}
\label{tab:comparison}
\end{table}
%Statistical methods are mainly based on probabilistic measures or other metrics such as frequencies or tf-idf scores.

A general overview of the NLP methods that are used in Android security is presented in Table \ref{tab:comparison}. We categorize the methods in four classes: rule-based, statistical, corpus-based, and neural models. Every method has its own advantages and disadvantages. For example, corpus-based models that use structured (e.g. WordNet) or unstructured (e.g. dictionaries) corpora  are quite robust in estimating semantic similarity; however, it is hard to find such available sources in many languages apart from English. The table shows that most of the studies in Android security use either statistical models such as tf-idf vectors, machine learning classifiers etc. Another majority of the studies utilize annotated dictionaries such as WordNet \citep{Wordnet}, or other manually built ontologies. The table shows that neural models are a new era for Android security, and only some of the recent work makes use of neural networks for security purposes. It is worth to note that in some studies, more than one of these categories might be utilized. Here, we categorize that study based on the dominant method. 

\subsection{Future Directions of the NLP Techniques in Android Security}

Semantic parsing has been a new emerging task in NLP in the last decade \citep{amr, abend, ucca}, which aims to learn the underlying semantic representations of text by finding the relations between linguistic utterances as in syntactic parsing, but this time learning the semantics. This could be a potential future direction in Android security as well, since most of the security-related problems require parsing the textual data also from the semantic perspective, such as finding the permissions in a description, finding the sensitive input entries data under privacy, or detecting the conditions in privacy policies which all refer to actions in a security-related scenario. 

The contextual word representations are the new state-of-art among the representation learning techniques \citep{word2vec, glove, fasttext}. BERT \citep{bert}, Elmo \citep{elmo}, Roberta \citep{roberta} are only few examples in this new trend. They have shown superior performance in many NLP tasks, such as machine translation \citep{mt-bert}, question answering \citep{qa-bert}, dependency parsing \citep{dp-bert} etc. We believe that the usage of deep contextualized word embeddings will also help in the tasks under Android security. For example, in the description-to-permission fidelity problem it is important to extract the correct meaning of the word \textit{contact}, which might be related to \textit{READ\_CONTACT} permission or not (e.g. in sentence ``Please contact us on this number.''), which can be only detected using the context of the word. In fact, it is worth to mention that BERT \citep{bert} has been recently used to identify opt-outs in privacy policies on websites \citep{Kumar2020} although this study is not in the scope of the Android security. On the other hand, such neural models come with their own vulnerabilities to  adversarial attacks. Adversarial examples on especially neural NLP models have been investigated in the last few years \citep{ebrahimi-etal-2018-hotflip, glockner-etal-2018-breaking,pruthi2019combating}.

Last but not least, attention networks also could be incorporated in the security-related tasks. Those attention mechanisms \citep{Sutskever, self-attention} have been proven to be very powerful in almost any kind of NLP problem (e.g. machine translation, question answering, text summarization etc.) by finding the important bits of the textual data when specialized in a given task. They could be also used in security-related tasks \citep{alecakir2021attention}. 

All of these methods and various tasks actually pursue a comprehension goal, which introduces the concept of Natural Language Understanding (NLP), rather than Natural Language Processing (NLP). Here, the aim is to gain the computers more like a human-level comprehension, rather than gaining just processing capabilities. We believe that Android security will also earn its share with the advancements in NLU, by comprehending the descriptions, privacy policies, and other relevant security data much better with those new techniques.

\section{Conclusion} 
The availability of useful textual data in Android markets encourages researchers to associate security behaviour of applications with such data using NLP-based techniques. This study reviews these NLP-based proposals in Android security and shows that such techniques could be useful and complementary to traditional security solutions. The proposals are analyzed in four categories (description-to-behaviour fidelity, description generation, privacy and malware detection) and summarized in Table \ref{tab:outline}. These studies are in their early stage and could be evaded by attackers by using more complex strategies. Therefore, these proposals need to be improved by using more advanced NLP techniques in the future, and even will need to  evolve towards natural language understanding. Although currently, semantics is incorporated in Android security restrictedly by applying certain tasks from NLP, the new emerging tasks such as semantic parsing in NLP and more advanced deep learning methods will help improving the current success of the frameworks in Android security. %Particularly, %\textcolor{blue}{BAKILMAYAN YORUM: Burcu buraya NLP'de yeni gelismeler nezdinde birsey soylenebilir mi? veya senin eklemek istedigin birseyler varsa} 
Therefore, we believe that we will see more combination of NLP-based techniques with conventinal methods for Android security in the near future.  
\label{sec:Conc}

\section*{Acknowledgements}
This study is supported by the Scientific and Technological Research Council of Turkey (TUBITAK-118E141). The authors would like to thank TUBITAK for its support.

\bibliographystyle{plainnat}
\bibliography{sample}  %%% Remove comment to use the external .bib file (using bibtex).

\end{document}